\documentclass[10pt,conference]{IEEEtran}
\usepackage{cite}
\usepackage{amsmath,amssymb,amsfonts}
\usepackage{algorithmic}
\usepackage{graphicx}
\usepackage{textcomp}
\usepackage{xcolor}
\usepackage[hyphens]{url}
\usepackage{fancyhdr}
\usepackage{hyperref}
\usepackage{array}
\usepackage{booktabs}
\usepackage{authblk}

\newcommand{\fixme}[1]{\textcolor{red}{(\textbf{Fix Me}) #1}}
\newcommand{\myparagraph}[1]{\noindent \textbf{#1}.}

\newcommand{\remove}[1]{}
\newcommand{\revision}[1]{\textcolor{red}{#1}}
\renewcommand{\revision}[1]{#1}

\newcommand{\takeaway}[2]{
\noindent \emph{Takeaway\enspace#1.} \emph{\uline{#2}} \\
}

\newcommand{\bryt}[0]{Bryt}

\begin{document}

\title{IPU: Flexible Hardware Introspection Units}

\author[1]{Ian McDougall}
\author[1]{Shayne Wadle}
\author[1]{Harish Babu Batchu}
\author[1]{Karthikeyan Sankaralingam}
\affil[1]{University of Wisconsin-Madison}


\maketitle
\pagestyle{plain}

\begin{abstract}
Modern chip designs are increasingly complex, making it difficult for developers to glean meaningful insights about hardware behavior while real workloads are running. Hardware introspection aims to solve this by enabling the hardware itself to observe and report on its internal operation — especially \textbf{in the field}, where the chip is executing real-world software and workloads. Three key problems are now imminent that hardware introspection can solve: A/B testing of hardware in the field, obfuscated hardware, and obfuscated software which prevents chip designers from gleaning insights on in the field behavior of their chips. To this end, the goal is to enable monitoring chip hardware behavior in the field, at real-time speeds with no slowdowns, with minimal power overheads, and thereby obtain insights on chip behavior and workloads. This paper implements the system architecture for and introduces the \textbf{I}ntrospection \textbf{P}rocessing \textbf{U}nit (IPU) - one solution to said goal. We perform case studies exemplifying the application of hardware introspection to the three problems through an IPU and implement an RTL level prototype. Across the case studies, we show that an IPU with area overhead $<1 \%$  at 7nm, and overall power consumption of $<25 mW$ is \textit{able to create previously inconceivable analysis:} evaluating instruction prefetchers in the field before deployment, creating per-instruction cycles stacks of arbitrary programs, and detailing fine-grained cycle-by-cycle utilization of hardware modules.

\end{abstract}

\section{Introduction}
\label{sec:intro}

Modern chip designs are increasingly complex, making it difficult for developers to glean meaningful insights about hardware behavior while real workloads are running. Hardware introspection aims to solve this by enabling the hardware itself to observe and report on its internal operation — especially \textbf{in the field}, where the chip is executing real-world workloads. Although prior work (performance counters, debug monitors etc. - see related work) has explored forms of introspection, success has been limited when confronted with three urgent challenges:
\underline{1. Lack of A/B Testing.} In modern software development, new features are tested side by side to collect performance data on live workloads \cite{feitelson2013development, fitzgerald2017continuous}. No equivalent mechanism exists for hardware. Architects cannot ``deploy'' or ``test'' a new feature in the field, observe real performance impacts, then decide whether to commit that feature to the next silicon revision. 
\underline{2. Obfuscated Hardware.} Software developers do not see essential microarchitectural details, since current tools (e.g., performance counters) provide only coarse insights and hide the nuances of actual on-chip events \cite{10.1145/3579371.3589058}.
\underline{3. Obfuscated Software.} Hardware designers rarely know how their designs behave under real-world software usage. Information from in-field use does not percolate back to chip designers. 
These issues have become especially acute because both hardware and software are expanding in complexity. Microarchitects must handle larger design spaces, while software developers have difficulty optimizing code for intricate, opaque hardware. The recently proposed Time-Proportional Event Analysis (TEA) module, which tracks per-instruction cycle stacks (PICS) to aid software optimization using a specialized hardware design~\cite{10.1145/3466752.3480058, 10.1145/3579371.3589058}, argues performance counters are insufficient.

In this paper, we revisit hardware introspection with a new approach designed to bridge these gaps comprehensively. We present a framework that not only offers fine-grained observability to software but also allows hardware designers to gather actionable insights \textbf{in the field}, ultimately informing the chip design cycle.

A central question for hardware introspection is: \textit{How can we design future chips so that developers can capture, analyze, and derive insights from cycle-level data in real deployments?} Addressing this involves four key challenges: 
i)  Capturing fine-grained hardware signals \textbf{in the field} without incurring prohibitive area or power costs;
ii) Enabling programmability so that what introspection is done can be changed at runtime, rather than being fixed in silicon;
iii)  Deciding what to introspect on; 
iv) Controlling the data volume so continuous introspection does not overwhelm the system.

\begin{figure*}
\includegraphics[width=\textwidth]{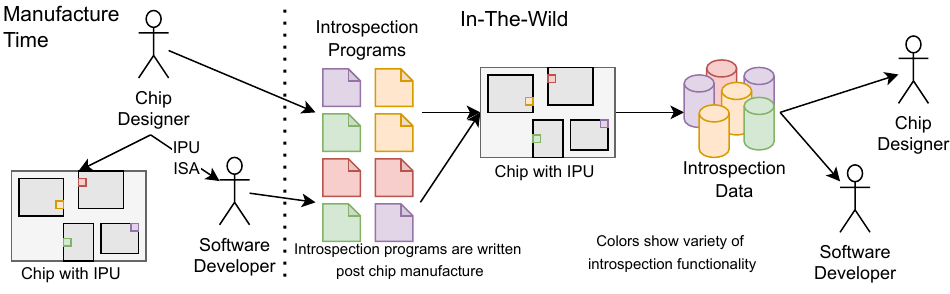}
\caption{IPU overview}\label{fig:ypuoverview}
\end{figure*}

Our solution, the Introspection Processing Unit (IPU), outlined in Figure~\ref{fig:ypuoverview}, tackles these challenges with a philosophy that balances efficiency and flexibility. Each IPU is built around a tiny RISC-V core placed in close physical proximity to the hardware block (a “Hardware-module Introspection Target” or HIT) whose signals it introspects. IPUs are integrated into HITs' physical hierarchy (similar to performance monitors) removing the need for costly wiring and, at design time, allowing engineers to profligately choose which signals are worth exposing - up to 32 signals per IPU. Chip designers, at design time, choose what the HIT is and attach an IPU to it - there can be multiple HITs and corresponding IPUs on a single chip. We expect HITs to be small - around 3 to 4 $mm^2$. Meanwhile, the actual analysis that runs on these signals is fully programmable, can be conceived of and implemented post-manufacturing any time during the chip's lifetime. From a system-level perspective, the IPU appears as a small PCIe device, providing a logical FIFO to send introspection outputs back to the host. This ensures that data volume remains manageable and that the host software stack can enforce bandwidth limits if needed. Chip designers can use software signing mechanisms to control 3rd party introspection for HITs. In this way, the IPU design forms a practical solution for hardware introspection delivering (i) efficient at-speed data capture, (ii) flexible analysis that allows rich hardware introspection, and (iii) low-overhead data handling. 


This paper's contributions are the definitions, system design, and implementation of the IPU, including an RTL implementation, that enables such a type of introspective hardware. To evaluate the IPU, we show 3 case studies using the IPU along the 3 key problems. \\
$\bullet$  \underline{A/B testing.} \revision{To address the lack of A/B testing for hardware, we demonstrate how the IPU can be used to evaluate a state-of-the-art entangled prefetcher~\cite{entangled-prefetcher}. Specifically, the IPU allows chip designers to emulate and evaluate a new prefetching algorithm on real workloads running in production systems—without modifying the hardware pipeline or incurring runtime slowdowns. This form of in-field evaluation enables comparative testing of candidate microarchitectural features. While fundamentally different from software A/B testing due to fabrication cost and rollback limitations, the IPU provides a practical mechanism for lightweight, post-silicon experimentation with hardware behaviors via introspection programs.}

    \noindent $\bullet$ \underline{Obfuscated hardware.} Recent work has shown that per instruction cycle stacks (PICS) provide extraordinary insights into hardware behavior and opportunities for software optimization, way beyond traditional performance~\cite{10.1145/3466752.3480058, 10.1145/3579371.3589058}. However the implementations require specialized hardware to implement PICS. We show that PICS can readily be implemented as a program that simply runs on our IPU, with inputs being easy to access signals of the processor's microarchitecture. \textit{In essence, we show that an IPU achieves the functionality of the specialized PICS hardware implementation while being programmable.}
    
    \noindent $\bullet$ \underline{Obfuscated software.} Conversely, hardware designers cannot observe enough about the software. In the case of GPUs this becomes quite acute. In spite of their extensive performance counters and profiling libraries, GPU designers have little in the field data about the software being run on their chips. We show that by collecting fine-grain cycle-level hardware utilization rates of key streaming multiprocessor (SM) components through an IPU, hardware developers can observe opportunities for overlapped execution.

In addition, the IPU is capable of providing the functionality of recent work like on-chip power estimation~\cite{10.1145/3466752.3480064}, historical works on data logging\cite{10.1109/MM.2009.6, 10.1145/1945023.1945034,10.1145/1005686.1005708}, monitoring engines\cite{7056071,6835922,244034,1207014,244034}, and specialized support for debugging and watchpoints~\cite{greathouse2012case} to enumerate a few.

Our results show the IPU executes each case study correctly, capturing the functionality of a specialized design. Second, it demonstrates IPUs achieve introspection capability infeasible with existing techniques. Third, it is efficient - the area overhead is $<1\%$ and power overhead is $<25mW$, corresponding to $<1\%$ in the worst-case of the IPU being always active (with something as simple as 10\% sampling, the overhead reduces by that factor). Our results span 15 SPECCPU benchmarks, 135 Champsim traces, and 21 Gemm shapes covering both CPU and GPU uses. The simulation testbed, IPU RTL, and introspection code will be released for others to build upon as an artifact.

\revision{It is important to clarify that the IPU architecture is not a collection of custom, per-use-case monitors, but a shared, programmable unit that runs different introspection binaries over the same general-purpose hardware. Across all three case studies, the same IPU variant can be used without modification. This programmability allows chip designers to explore hardware behavior in the field with significantly more flexibility than fixed-function PMUs. For example, our prefetcher emulation case study (\S\ref{sec:case-studies}) implements decision logic to test new algorithms on live workloads—functionality that is fundamentally infeasible with PMUs. While end users may not directly interact with IPUs, the design targets chip vendors, system integrators, and firmware developers who routinely require deep observability in controlled environments. The IPU provides them with a secure, low-overhead mechanism to introspect post-silicon behavior with new kinds of insights, ultimately informing the evolution of hardware design.}

This paper is organized as follows. An overview of the IPU and surrounding system is provided in \S~\ref{sec:design}, software and hardware architecture are detailed in \S~\ref{sec:software} and \S\ref{sec:hardware} respectively. Evaluation methodology is in \S~\ref{sec:methodology}, and our results with the three case studies are in \S~\ref{sec:case-studies}.  \S~\ref{sec:background} covers related work.

\section{System Overview}
\label{sec:design}


\begin{figure*}
    \includegraphics[width=\textwidth]{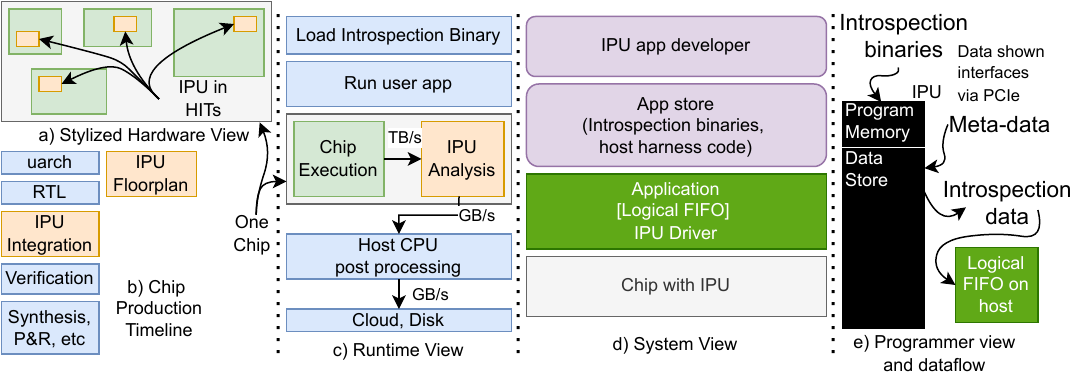}
    \caption{Cloud is purple, chip/HIT is green, blue is existing processes, orange is our contribution/new circuitry}
    \label{fig:info-flow}
    \label{fig:security-overview}
\end{figure*}


We provide a full overview of the software and hardware components of an IPU as shown in~\autoref{fig:info-flow} before providing the details in the following sections. To provide some context, we outline two types of introspection needed. First, for simplicity one that can be done in other ways as well. Consider the outputs of a Tensorcore every cycle. One introspection program is to build histograms (say 1024 equal-sized bins) of distributions of values as they are produced. The data producer would be Tensorcore, the signals are its output values, and the IPU would run instructions every cycle to determine the bins those values belong in. The histogram could be sent to the host's memory every 1 million cycles. Second, considering A/B testing - an introspection program could be instruction prefetch logic, with the data producer being a CPU's front-end block, with one signal - the {\tt current PC} being fetched. The introspection output is the effective miss-rate and accuracy. 

The programs run on the IPU that perform introspection we call ``introspection binaries''. The term ``user program'' refers to an application that runs on the chip - like web-browser, Photoshop, DL inference, etc. Introspection programs analyze user programs. 

\myparagraph{Hardware}
The Introspection Processing Unit (IPU) is a modular design built around a simple in-order RISC-V core. Inclusion of common analytics functions beneficial to maintain speed creates the IPU$_{lite}$. The IPU$_{pro}$ has soft logic for finite state machine traversal introspection programs as we detail in \S\ref{sec:hardware}. An IPU is responsible for monitoring an individual hardware component of the underlying chip: the Hardware-module Introspection Target (HIT). Our design associates one IPU per HIT (Figure~\ref{fig:info-flow}(a)). Examples of possible HITs include individual sub-modules within a core, L2 controller, GPU SM sub-modules, a CPU core etc. The IPU will be flattened into the hierarchy of the HIT for placement and routing (P\&R) purposes (Figure~\ref{fig:info-flow}(b)). The IPUs are also connected to the underlying chip's on chip network (OCN), which is used to transfer produced introspection information out of the IPU. Introspection programs output small packets at long intervals (e.g. 1 million cycles) so the generated traffic does not impact the user program's performance.

\myparagraph{Runtime View}
The runtime view of a program is cascaded in Figure~\ref{fig:info-flow}(c) with time flowing downward. The IPU driver exposes an API to configure and load the introspection code. The user application can specify what regions to analyze (e.g. specific tensor address regions). Once configuration is complete, the user application begins execution. On each data input to the IPU, it determines if the data is within the region to analyze. If so, the introspection program executes on said data. While analysis is on-going, any further incoming data is dropped. Therefore, introspection programmers must be cognizant of the data rate specifications from the hardware designer. Once analysis execution completes, the IPU awaits the next data. In most introspection programs, outputs occur at intervals through a logical FIFO (Figure~\ref{fig:info-flow}(e)). Once the user application ends, the IPU output can be post processed and, finally, exported onto disk or into the cloud.



\myparagraph{System Architecture} The system architecture (Figure~\ref{fig:info-flow}(d)) includes an API to expose IPUs and the IPUs themselves. At runtime, the IPU appears as a PCIe device programmable via a simple API, allowing hardware and software developers to download introspection programs. Each IPU is physically tied to its hardware block, and secure code execution is enforced via certificate-based code-signing using public/private key infrastructure, similar to Android/iOS app stores~\cite{androidsigning,iossigning}. This enables a secure, app-store-like model where trusted introspection programs—authored by chip designers, researchers, or developers—are deployed through configuration API calls.

\myparagraph{Privacy}
\label{sec:security}
Figure~\ref{fig:security-overview}(e) presents an abstraction to understand the IPU's privacy and security implications. An IPU, by design, cannot inject signals into the hardware HIT; it only accepts binaries and meta-data as inputs and emits introspection data. Integrity of binaries is addressed via code-signing. Richer program analysis techniques inherently risk leaking microarchitecture details—PICS~\cite{10.1145/3579371.3589058}, for instance, reveals bottlenecks that reflect design decisions—as earlier shown by Desikan et al.~\cite{10.1145/375212.379271} using performance counters. The central privacy question is what unintended inferences introspection might allow; for example, end-users learning undisclosed HIT details. To mitigate this, we propose three policy modes: \textbf{closed}, where no third-party introspection is allowed; \textbf{restrictive}, requiring source-code review before code-signing; and \textbf{permissive}, permitting introspection from credentialed developers. These policies are enforced during code signing, with designers leveraging HIT semantics for decision-making. Techniques such as program verification~\cite{bouncer}, trace wringing~\cite{trace_wringing}, local differential privacy~\cite{rappor,shuffle-dp,cormode2018privacy,ldp-tianhao,51929}, and secure multiparty computation~\cite{cryptoeprint:2020/300} offer potential for future formal guarantees. A subtler issue involves using HIT$_x$ introspection to infer data about unrelated HITs or chip-level behavior—akin to side-channel attacks or industrial espionage, and while possible, the IPU’s risk is comparable to existing interfaces like performance counters. Our system-level API ensures strong user privacy by streaming only introspection data, without host state access (e.g., IP or MAC addresses). Lastly, trust between chip vendors and users remains a broader concern not unique to IPUs; telemetry and diagnostics (e.g., performance counters, JTAG) are often disabled in untrusted settings. CPUs and GPUs support confidential modes that disable monitoring, as with NVIDIA Hopper’s secure execution mode~\cite{nvidiaHopperSecurity}, and Intel’s diagnostic firmware requires signing. Similarly, IPUs can support SKU variants or boot-time configurations that disable hardware or introspection paths entirely. In untrusted deployments, IPUs would be disabled, aligning with industry norms and not detracting from their value in trusted or OEM-controlled contexts.

\vspace{-0.15in}
\section{Software Architecture}
\label{sec:software}
This section describes the software implementation on top of IPUs with a deep dive on introspection programs. These programs, called ``introspection binaries'', include code that will be run on an IPU. The collection of IPUs on chip appear as a single PCIe device with distinct memory mapped areas for each IPU. The IPUs also expose a host API for configuring what introspection binary to run and the trigger logic. ``User program'' refers to an application that runs on the chip (CPU or GPU) - like web-browser, DL inference, etc.

\subsection{Programmer's Model}
An IPU is a programmable RISC-V core with a unique interface: 32 named inputs per execution. API calls configure which regions of the user program are analyzed. A small on-IPU memory is available, and the introspection program runs once per input set, repeating if new data arrives or idling otherwise. Post-processing can be performed at program end before offloading results.


\myparagraph{Data \& Transfer Semantics}
A HIT’s signal interface to these 32 named values would be exposed to users through a formal documentation like an ABI Spec specifying semantics of signals and data arrival rate. Introspection developers must ensure that the HIT signal production rate matches data processing speed in the IPUs, or use sampling, or be cognizant that some data will be dropped if there is a mismatch. 
Introspection binaries are able to transmit data via host-memory mapped into the IPU's address space. \remove{For example, on a PCIe device, a host-memory FIFO can be mapped to the device.} The IPU can then issue simple memory instructions to the unit's memory hierarchy that are then transparently routed to a region in host memory. Using this, we can create a store for introspection results.

\revision{One might ask why we don’t add input buffers to avoid dropping data when the IPU is busy. While small buffers delay overflow, they cannot prevent it if the introspection rate is slower than the data arrival rate—by Little’s Law, loss is inevitable without stalling the HIT, which our architecture prohibits. Thus, the IPU drops inputs when active, and introspection programs must be designed with this in mind, using sampling, aggregation, or exploiting event sparsity. Though the IPU runs at 2\,GHz in 7\,nm, HIT modules may run faster (e.g., 3--4\,GHz). We do not require clock synchronization; instead, we support three modes: (1) Accept reduced fidelity via sub-sampling (e.g., 1-in-2 cycles); (2) Use a fast buffer for asynchronous, windowed processing; or (3) Restrict introspection to low-frequency phases or optimize the IPU for speed.}

\subsection{Software API and Execution Management}
\label{sec:config}
For configuration, the IPU exposes a minimal host-side API which are facilitated via memory-mapped I/O to the IPU. To configure a binary for execution, \verb|IPU_CONFIG_IMAGE(image)| specifies that an IPU should load a given introspection binary \verb|image|. To configure when analysis begins, \verb|IPU_CONFIG_START(addr)| sets the IPU to begin processing new data when the program to be analyzed reaches \verb|addr|. \verb|IPU_CONFIG_STOP(addr)| similarly sets when to stop processing new data. To enable fine-grained analysis we expose \verb|IPU_PAUSE()| and \verb|IPU_RESUME()| to  pause and resume introspection execution. Finally, the command \verb|IPU_FINALIZE()| instructs the IPU to execute any clean-up code necessary for the introspection program.  The API resembles CUPTI~\cite{nvtx} or Linux's \texttt{perf\_event}~\cite{linux-perf}, allowing either a harness process to transparently configure and trigger IPU execution, or user binaries to use \texttt{PAUSE}/\texttt{RESUME} for fine-grained control.

\subsection{IPU Program Lifetime}
\revision{This section is an end-to-end example. The HIT is a CPU core and analytics code development is under a closed policy - the CPU core designer will also write the analytics code. The analysis is PICS generation; the Obfuscated Hardware case study (\S\ref{sec:case-study-PICS}) - we encourage the reader to skim that first.}

\myparagraph{Pre-Fabrication} \revision{As part of the design of the CPU core, up to 32 signals are chosen to connect to the IPU - based on important signals in the microarch pipeline. No information needs to be released to the public because of the closed policy; yet, an internal ABI Spec would be created to facilitate analytics code development. A partial ABI Spec is seen in Table \ref{tab:ABI-spec}. Prior to verification, the IPU is flattened into the core layout and the HIT-IPU connections are made as outlined in Figure \ref{fig:info-flow}. A subtle issue is that the HIT designer needs to determine what the important signals are - by providing up to 32 we give them freedom to be profligate to allow rich analytics post-manufacture.}

\begin{table}[]
\small
    \centering
    \begin{tabular}{c|c|c|c|c}
        Signal & \#Bits & Reg & Semantics & Rate \\
        \hline
        itlb-miss & 1 & x0 & Instruction TLB miss flag & 1 \\
        icache-miss & 1 & x1 & L1 Icache miss flag & 1 \\
        recycle & 1 & x9 & Recycle ROB unique IDs flag & 1 \\
        fetch-pc-head & 64 & x11 & Next PC to be fetched & 1 \\
        \multicolumn{4}{c}{...} 
    \end{tabular}
    \caption{Partial ABI Spec. Rate measured in cycles between data points.}
    \vspace{-0.2in}
    \label{tab:ABI-spec}
\end{table}

\myparagraph{Development} \revision{With the ABI Spec defined, the analytics code can be developed, which is the PICS generation in our example here. To this end, the CPU designer references the ABI Spec for each of the 17 signals necessary and identifies which input registers they are connected to. A portion of the code handling the instruction TLB miss event is shown below:}
\begin{verbatim}
_main: regtimer 50000, psv_loop
psv_loop: beq x0, 1, itlbm_m
beq x1, 1, icache_miss; x1 is a HW inp sig
...
itlb_m: hash r1, x12; x12 is an HW inp sig
ld r2, r1, 0
addi r2, r2, 0x40
st r2, r1, 0
ret
...
\end{verbatim}
\revision{In the development a 400,000 cycles sample rate is chosen to limit the output load; this does create approximation error which the CPU designer tests and finds it within acceptable limits - the overall PC ordering by most cycles used is correct. The CPU designer releases the analytics binary, output location, approximation error, and if a region of interest can be chosen onto the app-store.}

\myparagraph{Analysis} \revision{Now, a SW developer has encountered unexpected slowdowns in their application and wants to profile it. They can download the PICS generation binary on the host. The listing also indicates that output will be put in a file in disk and that the user can optionally specify a region of instructions to analyze. They include a few API calls at the top of their program source: }
\begin{verbatim}
    IPU_CONFIG_IMAGE("PICS-generation")
    IPU_CONFIG_START(ROI_BEGIN)
    IPU_CONFIG_STOP(ROI_END)
    <program code>
\end{verbatim}
\texttt{ROI\_BEGIN} and \texttt{ROI\_END} \revision{are the beginning and end of the region of interest where the developer believes the slowdown to be. Essentially it sets the respective Program Counters as the address to monitor for activating the IPU.}

\myparagraph{Post-Analysis} \revision{The output file has a list formatted as in Table~\ref{tab:PICS-example}. Produced by host code, monitoring the IPU's introspection program. The developer uses the results for application performance tuning.}

\begin{table}[]
    \centering
    \begin{tabular}{c|c|c}
        PC & Event Combination & Number of Cycles  \\
        \hline
        0x7912d0 & DTLB miss, DCache Miss & 50000000 \\
        0x80dda0 & Branch Mispredict & 200000 \\
        \multicolumn{3}{c}{...}
    \end{tabular}
    \caption{A few rows of the output file from PICS generation analysis.}
    \label{tab:PICS-example}
\end{table}

\vspace{-0.05in}
\section{Hardware Architecture}
\label{sec:hardware}


\label{sec:ypu_arch}
\begin{figure*}
\centering
    \includegraphics[width=0.8\textwidth]{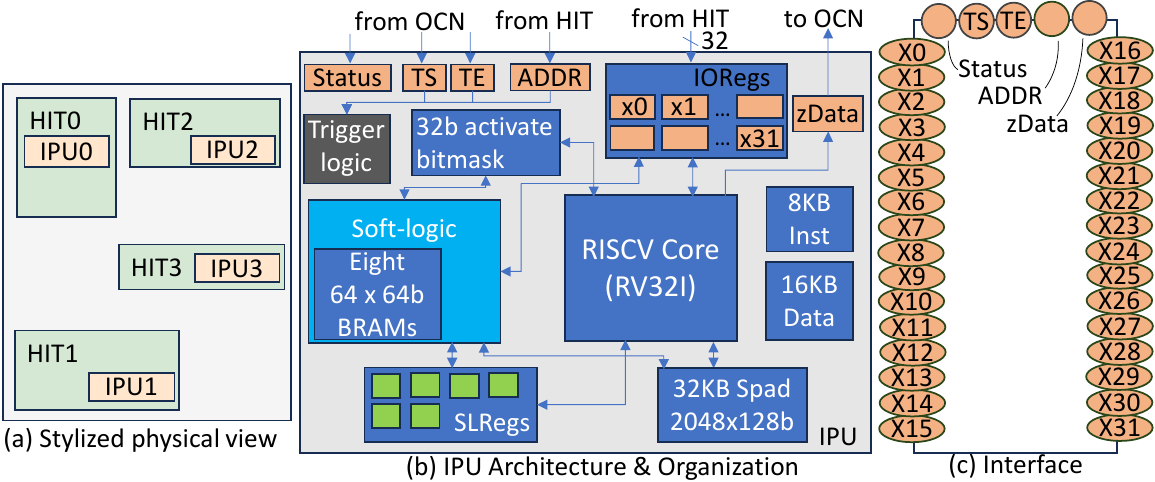}
    \caption{IPU hardware architecture.}
    \label{fig:detailed-arch}
\end{figure*}

A chip can have multiple IPUs. Each IPU observes hardware signals from its corresponding HIT and runs introspection binaries on those signals as stylized in Figure~\ref{fig:detailed-arch}(a). The IPUs use the chip's OCN, which enables them to transmit introspection outputs (which are infrequent) to and configuration from the host. The collection of IPUs is visible as a PCIe device and uses PCIe to interface with the host; each IPU is distinguished through distinct memory mapped regions.
If the chip lacks an OCN, some form of network which connects the IPUs to the PCIe interface must be added as well. 

Configuration of an IPU occurs through an API call:\\ \verb|IPU_CONFIG_IMAGE(image)|. This bundles the introspection binary (and trigger logic meta-data - details in \S\ref{sec:hardware}) and sends it to the appropriate IPU based on a given hardware device ID.

\myparagraph{Interface overheads} HIT-IPU connections are short because the IPU is flattened into the HIT during P\&R. Therefore, wiring overheads are negligible. Depending on the signals in a HIT and what the HIT itself is - there could be timing issues that can be addressed with standard buffering techniques used for performance counters. Consider a square HIT that is 2$mm^2$. At a simple level, signals might need to traverse $2.8mm$ (half the perimeter of the HIT) to reach the IPU logic. To avoid timing issues for a high frequency design, one flip-flop might be necessary. Since this ``far away'' signal is buffered by one cycle, it implies all signals of this $HIT \rightarrow IPU$ must be buffered. HIT designers can use P\&R feedback to judiciously select signals to avoid/minimize this.


There is no type of cross-chip wires. IPU configuration and output transfers occur over the chip's OCN and PCIe; both transmit small packets at long intervals meaning the traffic is negligible. We acknowledge, that even this meagre PCIe introspection traffic can introduce QoS and interference leading to non-linear slowdowns. Optimizations to this traffic management are future work.


 
\label{sec:sie_arch}
\begin{figure*}
    \includegraphics[width=\textwidth]{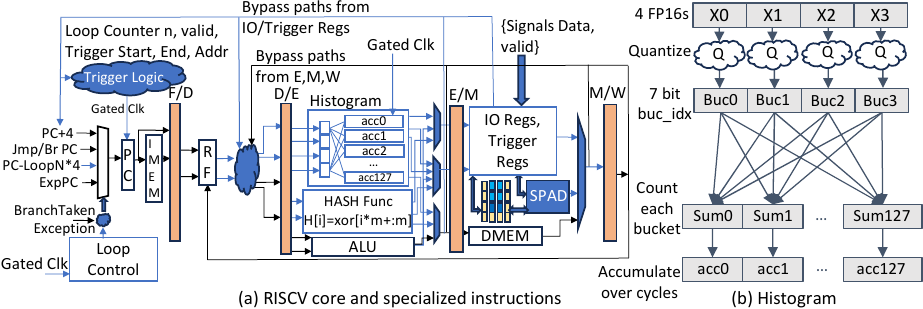}
    \caption{IPU Microarchitecture showing datapath and control-path changes}\label{fig:sie-uarch}
\end{figure*}

\myparagraph{Hardware organization}
The IPU architecture comprises of four baseline components:
a programmable core, a scratchpad SRAM, 32 input registers (IORegs), and 3 trigger registers: Trigger Start (TS), Trigger End (TE), and ADDR. The Trigger logic looks at the ADDR signal from the HIT along with TS and TE (programmed using API calls) to control when the IPU becomes active. This is shown in Figure \ref{fig:detailed-arch}(b) while the interface is laid out in Figure \ref{fig:detailed-arch}(c).
To concretize the architecture, we select some sizes for these components: the core is 32-bit RISCV core (RV32I instruction set) using a data-SRAM and an instruction-SRAM;
the scratchpad is 32KB.

\myparagraph{Execution Model}
An IPU has a 4-bit STATUS register, putting it in 6 states: PAUSED (P), ACTIVE-PAUSED (AP), ACTIVE-RUNNING (AR), FINALIZE (F), ERROR (E), and UNDEFINED (U). Its program structure includes 3 predefined program regions: {\tt init, \_main}, and {\tt end}. Borrowing from the simplicity of micro-controllers, {\tt init} is hardcoded to instruction memory address 0x0. The entire instruction memory is 8KB which amounts to 2048 instructions long. On power-on, PC is set to 0 and starts executing the code in {\tt init}. By convention, {\tt finish} is hard-coded to be 16 instructions from the bottom of the instruction memory at 0x7F0. When the IPU is set to the finalize state, it executes code in the {\tt finish} function and transitions to the PAUSED state.

The execution model of code on an IPU is data-driven, i.e. when new inputs arrive the {\tt \_main} function is called if the IPU is in the ACTIVE-PAUSED state. If it is running code triggered by previous input, it will be in the ACTIVE-RUNNING state - data received when in this state is dropped. Whenever we show a datapath of X bits for the IO registers, there is an implied additional valid bit associated. This bit is used by to determine whether new input has arrived. 

\begin{figure}
    \includegraphics[width=\columnwidth]{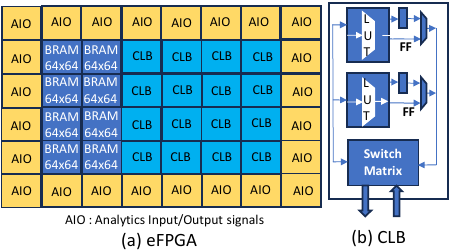}
    \caption{IPU$_{pro}$ soft-logic design}\label{fig:ypupro-uarch}
    \vspace{-15pt}
\end{figure}

\myparagraph{Microarchitecture} We design two IPU variants for different regions of the analysis-vs-data-rate design space. \textbf{IPU$_{lite}$} is a compact, cacheless RISC-V core with built-in primitives like histograms, loop counters, and hash functions for efficient, low-complexity introspection. \textbf{IPU$_{pro}$} augments the RISC-V core with soft-logic—a lightweight embedded FPGA with configurable logic blocks and small BRAMs—enabling complex, high-throughput analysis tailored at runtime. It interfaces via memory-mapped registers and supports introspection programs that bundle RTL logic alongside control code. Each IPU is embedded with its associated HIT and communicates via the OCN, avoiding long wires and limiting system traffic due to low data rates. Multiple IPUs scale well: 5 IPU$_{pro}$ and 10 IPU$_{lite}$ consume just 0.65\% of a 200\,mm$^2$ die, and even full coverage across GPU SMs stays under 1\% chip area overhead.

\section{Evaluation Methodology}\label{sec:methodology}
To empirically evaluate the IPU, we conduct three case studies that we briefly describe in the introduction. Each demonstrates the IPU's ability to resolve 1 of the 3 problems: \textbf{in the field A/B testing}, \textbf{obfuscated hardware}, and \textbf{obfuscated software}. Table \ref{tab:method-config} describes the emulation and simulation testbeds we built.

\begin{table*}

\raisebox{\dimexpr0.7\baselineskip-0.55\height}{\includegraphics[scale=1]{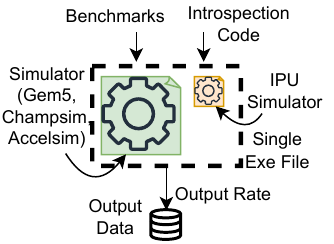}}
\begin{tabular}{c|c|c|c}
    Case Study & In the field A/B Testing & Obfuscated HW & Obfuscated SW \\
    \hline
    Benchmarks & 135 Traces \cite{CVP2} & 15 Spec17 & 21 Gemms \\ 
    \hline
    Simulator & Champsim & Gem5 SE & AccelSim \\ 
    Config Matching & Entangled Prefetcher~\cite{entangled-prefetcher} & TEA~\cite{10.1145/3579371.3589058} & QV100 Model \\ 
    \hline
    \#lines of IPU code & 300 Verilog & 75 & 2 \\ 
    \#bits into IPU & 64 & 215 & 4 \\ 
    \#signals from HIT & 2 & 17 & 3 \\ 
    \hline
    Output size & 3B  & 6B  & 4B \\ 
    Output timing (cycles) & Program & 400k &  256 \\ 
    Rate per HIT (1/s) & approx 0 & 15KB & 15.6MB \\ 
\end{tabular}
\caption{Methodology configurations and IPU System Testbed Flow for emulation in figure.}
    \label{tab:method-config}
\end{table*}



\myparagraph{Emulation and Simulation testbed} Our three case studies span prefetch engine (Champsim~\cite{Gober2022TheCS} simulator), core-microarchitecture (GEM5~\cite{10.1145/2024716.2024718} cycle level simulation), and GPU cycle-level simulation (AccelSim~\cite{9138922}). 
We built an IPU emulator for code development and to determine correctness of our introspection. For performance (time), we developed a co-simulation environment that adds an IPU simulator into Champsim, Gem5, and Accelsim (left Figure in \autoref{tab:method-config}). For area and power, we implemented RTL (and then synthesized) which was verified with an emulator for correctness. Table~\ref{tab:method-config} also shows the number of lines of code for the {\tt \_main} function of each introspection program. Overall, we have more than 171 applications simulated. 

\myparagraph{RTL Implementation}
We implemented IPU$_{pro}$ and IPU$_{lite}$ in Verilog. Our implementation was verified for many input values against the introspection reference implementation. We use the AsAP7 7nm educational PDK~\cite{CLARK2016105}. For SRAMs we use CACTI scaled from 32nm to 7nm per~\cite{stillmaker_scaling_2017}. We implemented our soft-logic using the FABulous design flow~ \cite{10.1145/3431920.3439302} to estimate area and power, and their synthesis flow for utilization. To determine soft-logic power, we used data from the reference introspection execution to create input traces. We used ASIC process flow of synthesis (DC Compiler/Primetime), APR(Innovus), and VCD based power estimation obtained from Netlist simulation of all case studies. The max clock frequency for the soft-logic and IPU$_{pro}$ is 1.3 GHz and 2 GHz for the RISC-V core i.e. IPU$_{lite}$. The HIT signals we need are described in the case studies and we show that the signals are readily available for any reasonable implementation of a GPU or core. The first two case studies are on a CPU and the third case study is on a GPU. To do comparison, our references are: a CPU Zen2 4-Core Complex~\cite{9063113} which needs 31 mm$^2$ area and consumes 4 watts of power~\cite{cpu-power} and a GPU SM that uses 3.475 mm$^2$~\cite{locuza-a100} of area and consumes 1 watt of power~\cite{gpu-power, kandiah2021accelwattch}.

\section{Case studies}\label{sec:case-studies}
\begin{figure*}
    \centering
    \includegraphics[width=\textwidth]{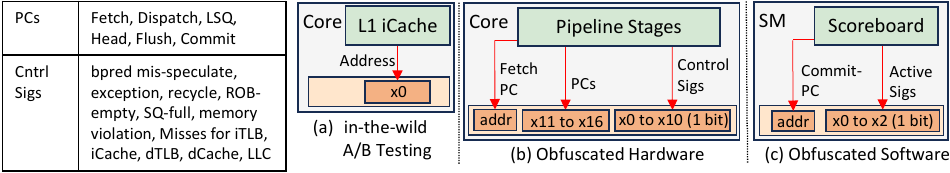}
    \caption{Case studies interfaces. (b)PCs and Control Sigs for PICS listed in left-hand table. (c)Active Sigs are for the tensor core, SIMT, and memory subsystem. }
    \label{fig:case-studies-overview}
    \label{fig:case-studies-pipeline}
\end{figure*}

\begin{figure*}[tbp]
    \centering\includegraphics[width=\textwidth]{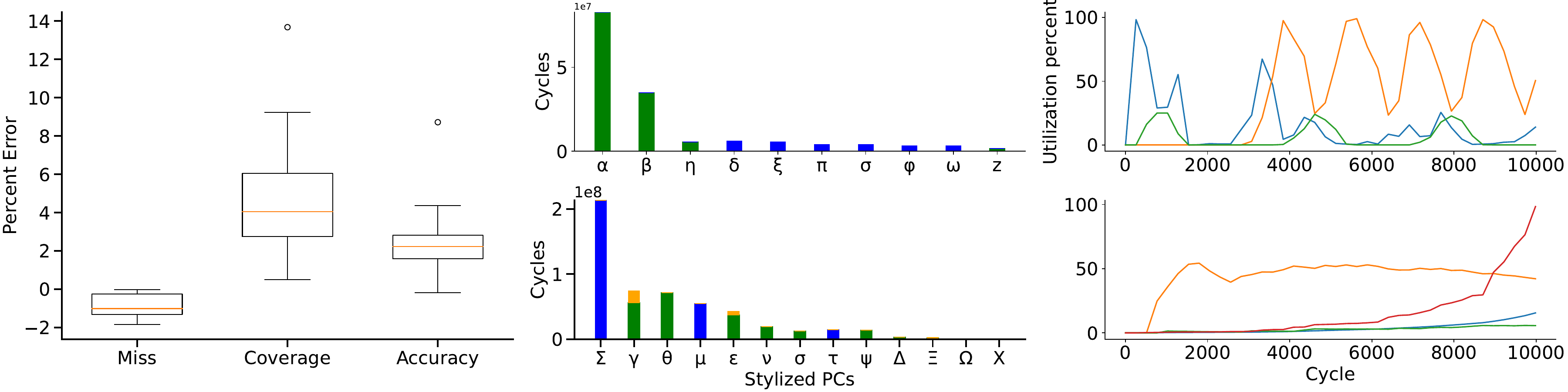} 
    \begin{tabular}{m{0.35\textwidth}m{0.3\textwidth}m{0.2\textwidth}}
     (a) In the field A/B Testing & (b) Obfuscated Hardware & (c) Obfuscated Software \\
    \multicolumn{3}{c}{In the interest of space, the labels for (b) are symbols. In reality they are PC values in the application binary.}
    \end{tabular}
    
    \caption{Utility results of the three case studies. 
    (a) Relative error for each metric in prefetch emulation in-silicon across 135 workload traces. 
    (b) TOP-10 PICS for NAB and Libquantum benchmarks, showing instruction contributions to exposed cycles. \fcolorbox{black}{green}{}~dcache miss, \fcolorbox{black}{blue}{}~Drain- SQ full, \fcolorbox{black}{orange}{}~Icache miss and Dcache miss 
    (c) Cycle-level GPU utilization. \fcolorbox{black}{cyan}{} SIMT,  \fcolorbox{black}{orange}{} Tensor Core, \fcolorbox{black}{green}{} higher level memory, \fcolorbox{black}{red}{} SIMT sorted by utilization. 
    }
    \label{fig:utility-cs1}
    \label{fig:utility-cs2}    
    \label{fig:utility-cs3}    
    \label{fig:utility-cs4}    
\end{figure*}

We now describe three case studies resolving the 3 key problems while spanning CPU and GPU and covering different HITs and signal types. \textbf{We emphasize that an IPU accurately captures each case study: enabling A/B testing that is impossible currently, constructing disruptive PICS stacks without dedicated single-function hardware, and gathering in the field characteristics of GPU utilization}. 
Figure~\ref{fig:case-studies-overview} shows which hardware signals are connected to the IPU. Figure \ref{fig:utility-cs1} depicts a utility result for each case study.
For each case study, we cover the key problem, how our use is exemplary of the problem, what utility the introspection holds,  and where applicable present a quantitative comparison, the design of the introspection code, the area and power, and whether any data is dropped and its effects on accuracy if so. An overview of the interface width, code length, and output characteristics is shown in Table~\ref{tab:method-config}. For context, PCIe bandwidth for A100 is 32 GB/sec~\cite{a100-in-depth}; our introspection output never exceeds 2.0 GB/sec.


\revision{This reflects worst-case raw output bandwidth under sustained introspection. In practice, IPU outputs are small and sparse—our case studies show useful analytics from just a few bytes every hundreds of thousands of cycles. Results are meant for in-system use: the GPU case study emits short utilization histograms, not full traces; the PICS study sends only aggregated delay signatures. Further aggregation or selective export would be done locally, avoiding raw trace streaming to the cloud.}

\vspace{-0.1in}
\subsection{In the field A/B Testing}
\begin{figure}
    \centering
    \includegraphics[width=0.8\columnwidth]{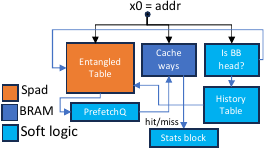}
    \caption{Organization of the soft-logic block for prefetcher}
    \label{fig:prefetch-layout}
\vspace{-0.1in}
\end{figure}

A chip designer cannot directly compare the effectiveness of hardware designs through deployment tests; this is unlike some software that deploys multiple designs and analyzes each for their effectiveness in real-world situations. It would be a hazard to have applications run on designs under test, so a tool to emulate the design's effectiveness while execution continues using the hardware made at manufacture time is ideal: this is complex analysis. To perform an effective test, all inputs should be captured meaning a high data arrival rate. Thus, the IPU$_{pro}$ is utilized.

This case study tests the recently proposed entangled prefetcher~\cite{entangled-prefetcher} via emulating a form of the prefetcher in the IPU$_{pro}$'s soft logic. In one core complex, one such IPU$_{pro}$ can be integrated to introspect the front-end of a core. The analysis produces the coverage, accuracy, and miss-rate for in the field applications. We achieve results within 1.8\%, average of 1\%, for miss rate of tests using simulations. 
\textbf{This demonstrates that the IPU can test hardware designs at speed in the field before fabrication enabling unprecedented analysis of new designs.}


\remove{\myparagraph{SOTA-Capability} This use case fundamentally requires cycle-level processing of values beyond distribution / aggregation. Thus, all four attributes of the taxonomy are necessary to support A/B testing of novel hardware designs. \revision{The current methodology collects a trace through software profiling then HW emulates or RTL simulates the component design. No single general approach can capture the required signal(s) and faithfully execute the design.} The LBA paradigm \cite{10.1145/1945023.1945034, 10.1109/MM.2009.6} is the closest amenable approach. However, none of the existing literature provides this capability out-of-the-box.} 


\myparagraph{HIT \& Interface} The HIT is the CPU front end block, and we need essentially one data signal - fetch PC being issued by the processor core (depending on the decoupled front-end design, the signal could be different; \revision{a virtual or physical address depending on cache design}). The analysis is performed over the entire program so no signal is connected to ADDR nor are TS and TE configured.

\myparagraph{Introspection Code} Implementing the prefetcher using RISCV code is too slow resulting in many dropped data (a state-machine traversal is heavily branchy code). Instead, we implement it on the soft logic and are able to run at one address per cycle with nearly full eFPGA utilization. The RTL design of the prefetch emulator is shown in Figure~\ref{fig:prefetch-layout}. Since we cannot inject anything into the HIT, our entangling design assumes that all L1 misses are L2 hits to determine entangling pairs (we measure the error this introduces). The execution of memory requests is still correct under this assumption - it does not change the contents of L1.

\myparagraph{Performance analysis} The performance analysis here is simple - the design accumulates coverage, misses, and accuracy counters in hardware and emits them periodically (every $2^{31}$ cycles to the host) to avoid overflow. Thus, the traffic to host is minimal.

\myparagraph{Simulation methodology and results}
In our Champsim testbed we ran 135 CVP2 traces and compared IPU based prefetch to the original entangled prefetch implementation from the authors. Our results are nearly identical to the results from the original paper, as the only difference is the always-hit-in-L2 assumption, discussed below. To measure error, we compare our statistics to the reference simulation's statistics.

\myparagraph{Analysis of approximations} Figure~\ref{fig:utility-cs4}(a) shows coverage, accuracy, and miss-rate error across the traces. For each metric, our always-hit-in-L2 assumption leads to better prefetching than actual, outperforming on each statistic by less than 5\% on average. 
In cases where the prefetch stats are high, the initial miss rate was very low (less than 0.25\%) so the other prefetch stats are less meaningful. The Figure shows the distribution of errors in terms in min, max, and inner quartile. \revision{Note that we don't model cache pollution effects, which our results show has small impact on accuracy.}

\myparagraph{Area and Power} The area of an IPU$_{pro}$ is 0.22 mm$^2$. In comparison to the CPU reference, this is 0.7\% area overhead; power is 20.8 mW, which  is around 0.5\% of the CPU reference (Section~\ref{sec:methodology}). On a chip with a single IPU$_{pro}$ instead of one per core complex, the area and power overheads reduce to 0.175\% and 0.125\% respectively.


\textit{Takeaway 1. An IPU enables in-field A/B testing of policies on real workloads, allowing behavior inference before silicon redesign—previously infeasible for microprocessors.}

\vspace{-0.1in}
\subsection{Obfuscated Hardware}\label{sec:case-study-PICS}

Modern hardware hides much of its internal complexity, leaving developers with limited visibility into how their programs actually execute. Performance counters have proven insufficient to bridge this gap~\cite{10.1145/3579371.3589058}. Per-instruction cycle stacks (PICS) reveal which static instructions dominate execution time and what core events occur during each dynamic instance, enabling significant speedups\footnote{Due to space limitations, we refer readers to the original TEA/PICS paper for design details~\cite{10.1145/3579371.3589058}. Our aim here is to match the TEA behavior; the original paper already validates its utility, which we re-verified using our emulation testbed.}. This compute-intensive analysis demands high-rate signal access, making IPU$_\text{lite}$—placed per core complex—a natural fit. Unlike prior work that used dedicated RTL~\cite{10.1145/3579371.3589058}, IPU$_\text{lite}$ constructs these stacks using its programmable core as illustrated in Figure~\ref{fig:utility-cs1}b.

\myparagraph{HIT  \& Interface} The HIT is the core pipeline of a CPU. Figure \ref{fig:case-studies-overview}(b) shows the interface as listed in the left table including Control Sigs, which indicate long-latency events starting in the core. In addition, we have 6 \revision{virtual address} PC values from 4 parts of the pipeline.
To restrict program regions, users set the {\tt TS} and {\tt TE} registers with the fetch-PC connected to the {\tt ADDR} register using the IPU API.

\myparagraph{Introspection code} The introspection code has two phases: every cycle we update a Performance Signature Vector (PSV) which is a bit-mask that indicates which event has occurred for a particular dynamic instance of a PC. This follows a sequential if-else-if sequence across all supported hardware events, where if an event occurs in the core pipeline a load-modify-store sequence sets a specific bit of the appropriate PSV to a '1'. If an instruction is flushed, we store its associated PC value in the IPU's memory, so that we can reference it after it commits. Every 400,000 cycles (TEA paper's design value) we update PICS which correspond to combining the delays of every dynamic instance of a PC into a single entry. The introspection code scans through the active list of PSVs to determine to which PSV we can attribute cycles and sends to the FIFO a payload comprising of PSV (PC + signature).

\begin{figure}
    \centering
    \includegraphics[width=\columnwidth]{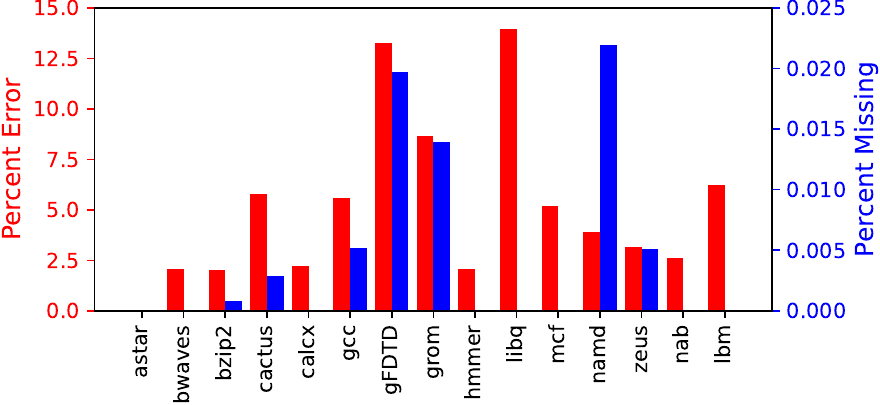}
    \caption{Average of the relative error for the PICS per benchmark in red and percent of total delay missed in blue}
    \label{fig:TEA-error}
\end{figure}

\myparagraph{Performance analysis}
215 bits of data are used every cycle in this case study. In the common case (representing more than 75\% of the cycles - our results and traces we obtained from TEA authors confirm this), no event is triggered when the ROB is sampled (as it isn't stalled/drained). In the 25\% eventful cycles, typically a single long latency event occurs. Some times 2 or more events occur when the ROB is sampled (very rarely 3 events, and almost never more than that). Outside of the cycles where the ROB is sampled, there are typically 1-3 events in the processor pipeline which need to be processed, triggering 3-9 instructions of code. The introspection output data volume  a few bytes of PSV data every 400,000 cycles.

\myparagraph{Simulation methodology and results}
As shown in Table \ref{tab:method-config}, we use a gem5-based simulation. For the SPEC benchmarks, we ran 1 billion cycles of simulation after fast-forward 1 billion cycles (matching~\cite{10.1145/3579371.3589058}'s  methodology). Two example PICS stacks from our 15 applications (all of which we generated) are shown in Figure~\ref{fig:utility-cs1}(b). For validation of PICS generation, we ran 3 DARCHR microbenchmarks~\cite{microbench_github} expecting \textit{one} PC to show up in the PICS stack for these microbenchmarks. The resultant PCs are shown below verifying the generation. \revision{One PC has a large cycle-count, showing that is primarily responsible for performance stalls.}

\begin{table}[h]
\footnotesize
    \centering
    \begin{tabular}{c|c|c|c} \hline
    PC & Assembly  & kCycles & C Code Line \\ \hline
    \multicolumn{4}{c}{STL2 causes LSQ Full} \\ \hline
    4017f6 & mov \%eax,(\%rsi,\%rdx,1) &127878&arr[lfsr].p1=lfsr\\ \hline
    \multicolumn{4}{c}{CCH\_st causes Branch Misspeculation} \\ \hline
    401813 & jne 4017f8 &177&if(randArr[i])\\ \hline
    \multicolumn{4}{c}{ML2 causes D-Cache Miss} \\ \hline
    4017ee & mov \%eax,(\%rsi,\%rdx,1) &58595&lfsr = lfsr +arr[lfsr].p1\\ \hline
    \end{tabular}
\end{table}

\myparagraph{Analysis of approximations}
This case study has the notion of dropped data - if an event is triggered during the PSV generation window of a previous event, we drop that event. Note that IORegs are designed to hold their ``old'' data (and drop new data) until the IPU reverts back to AP state. To understand the impact of this, we used our simulation testbed to create PICS with simulating introspection code running in 1 cycle vs per-cycle simulation of the introspection (which can take 8 cycles when two events occur in the same cycle). Our error metric is defined as average relative error (compared to the single-cycle version) of the cycle stack height for each PC for each application. Figure~\ref{fig:TEA-error} shows this in the red bars. Typically the quantitative error is $<3\%$ while a couple of applications show up to 12\% error. In all scenarios the list of PCs and the scale of the cycles contributions to PICS was correct, which is most important for performance optimizations.
In some very rare scenarios, we drop entire PCs from the PICS stack - when a PC always appears in the dropped window. The Y-axis shows the percentage of cycles covered by these dropped PC in blue. By definition these are exceedingly rare and unimportant for performance analysis. Across our applications, they cover $<0.025\%$ of cycles.

\myparagraph{Area and Power} The IPU$_{lite}$ has an area of 0.019 mm$^2$. Compared to the CPU reference, this is an area overhead of 0.06\%. Power consumption is 15.0 mW which is 0.38\% of the CPU reference power. On a chip with a single IPU$_{lite}$ instead of one per core complex, the area and power overheads reduce to 0.015\% and 0.095\% respectively.  


\textit{Takeaway 2. A IPU accurately captures PICS stacks showcasing it address the HW obfuscation problem.}


\begin{figure}
    \centering
    \includegraphics[width=\columnwidth]{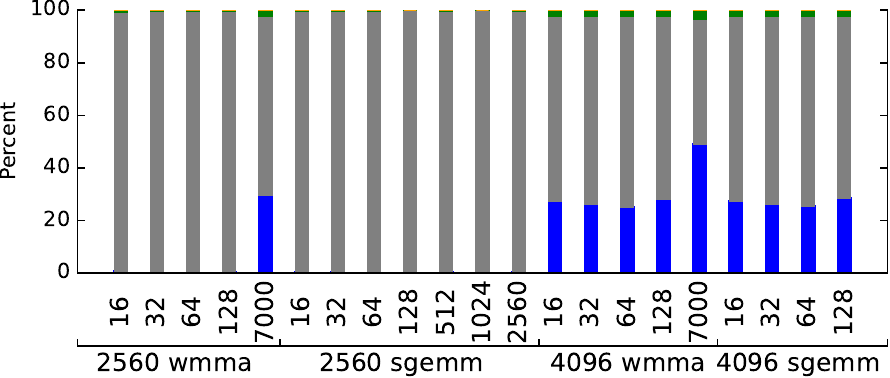}
    \caption{Breakdown of how use overlaps amongst the three signals collected shown per gemm shape. Blue is all low. Grey is 1 high and 2 low. Green is 2 high and 1 low. The number under each stack is the n and number in groupings is the m and k. wmma or sgemm are the kernels. The leftmost stack is (2560,16,2560) wmma benchmark.}
    \label{fig:gpu-util-ranges}
    \vspace{-0.2in}
\end{figure}

\subsection{Obfuscated Software}

Hardware designers rarely see how real applications behave post-fabrication, and benchmark suites quickly become outdated—especially with fast-evolving ML workloads—leading to a growing mismatch with in-field behavior. In the context of GPUs, as they evolve~\cite{h100-in-depth}, designers would benefit from cycle-level visibility, which motivates using IPU$_\text{lite}$ to capture high-rate, low-complexity insights. \textbf{IPU$_\text{lite}$ can generate histograms of fine-grained, cycle-level activity, revealing mutually exclusive usage patterns.} In optimized CUTLASS GEMM kernels, SIMT Core, TensorCore, and Memory Engine operate in largely disjoint cycles, highlighting a clear opportunity for overlap and performance gains without extra hardware or bandwidth.

\myparagraph{HIT \& Interface} This is a GPU case study, with the HIT for the IPU being an SM scoreboard block. The data signals are 3 one-bit signals indicating whether the SIMT core is active, the TC is active, and the L1 cache subsystem is in a state where it is servicing one or more outstanding requests (MSHRs non-empty status). While GPU hardware is proprietary the performance counters from NVIDIA Nsight Compute CLI (NCU) count aggregates for these signals indicating they are readily available and not disrputive from a design standpoint.
Optionally, \revision{the virtual address} retiring PC is connected to the ADDR register with IPU's software API used to restrict regions of interest. To isolate to a region within a kernel, PC start and end range be provided to TS and TE. Or it can be run without trigger to capture this activity for the entire kernel.

\myparagraph{Introspection code}
The introspection code is a single histogram instruction (optimized with a loop directive) that runs for 256 cycles receiving new data every cycle. The output is 3 bytes every 256 cycles, denoting how many active cycles of that unit, which can also be batched across windows.


\myparagraph{Performance analysis} Every 256 cycles, we emit three 1-byte values; thus, introspection output data bandwidth is 
3 * 108(\#SMs) / 256 bytes per cycle = 1.7 GB/second at 1.4 GHz. This frequency of this host traffic can be further reduced by batching in the IPU's data-store. By using longer windows the traffic can be further reduced.

\myparagraph{Simulation methodology and results}
One representative output is shown in Figure~\ref{fig:utility-cs3}(c). 
The top half shows chronologically ordered windows of 256 cycles with the Y-axis denoting \% of cycles in that window where that signal was active. We can see the mutually exclusive behavior. The bottom graph shows the same data in a histogram format: the windows are sorted in increasing order of utilization, and we plot the running average up to that window for the signals. Figure~\ref{fig:gpu-util-ranges} post processes this data and presents it in a different way. We classify windows of 256-cycles into 4 bins: 2 signals high (green), one signals high (blue), all signals low (grey). Where high mean greater than 25\% the cycles in that window, and low meaning less than 25\% of the cycles. We can see that large portions of time are spent with at least one component of the SM being idle - pointing to further hardware optimization beyond directions like TMA~\cite{h100-in-depth} that have appeared. Other work has also looked at improving such utilization ~\cite{gpu-parallelism, gpu-utilization-combo}.

\myparagraph{Analysis of approximations}
This case study uses our histogram instruction in a novel way, essentially treating each signal as its own bucket and builds a 3-bucket histogram. Hence we have very little dropped data - the source of error is the few cycles needed at the end of a window to write 3 bytes of accumulated statistics.

\myparagraph{Area and Power} The IPU$_{lite}$ has an area of 0.019 mm$^2$ that is 0.6\% of the area of the GPU reference. This case study consumes 4.7 mW of power as determined by Section~\ref{sec:methodology} methods. The IPU$_{lite}$ power overhead is around 0.5\% of GPU reference. If instead the chip designer included only one IPU$_{lite}$ on the chip, the area and power overheads are 0.003\% and 0.004\% respectively.


\textit{Takeaway 3. An IPU enables fine-grained, simulation-level GPU analysis in the field, revealing optimization opportunities in concurrent resource usage.}

\subsection{Comparison with Existing Mechanisms}\label{subsec:case-study-SOTA-comparison}

\revision{\textit{Across all three case studies, the IPU enables runtime, signal-level introspection beyond the reach of fixed-function hardware, tracing tools, or postmortem analysis.} We summarize each case study’s relation to existing tools. In the first, the IPU emulates a candidate prefetcher via custom logic triggered on memory accesses—something not possible with commercial tools like Intel PCM, CUPTI, or CoreSight, which lack in-field programmable logic. While LBA~\cite{10.1145/1945023.1945034, 10.1109/MM.2009.6} is conceptually closest, it doesn’t support user-defined prefetch emulation in production. In the TEA case, performance counters cannot implement Time-Proportional Event Analysis (TEA)~\cite{10.1145/3579371.3589058}, which underpins PICS. Unlike prior work requiring dedicated RTL for RISC-V BOOM, our IPU replicates TEA without RTL changes and with greater flexibility. Finally, the IPU builds histograms of fine-grained, cycle-level signal states, which are impractical to obtain via simulation, emulation, or debug monitors due to visibility limits or high overheads (e.g., 100$\times$ slowdown). Profiling co-processors~\cite{903267} offer partial alternatives but struggle with non-uniform, fine-grained utilization.}

\section{Related Work}
\label{sec:background}

We define a four-attribute taxonomy to contextualize the IPU in related work (\autoref{tab:primitives-other-soln}). \textbf{S}peed measures how quickly introspection can run without disrupting execution; \textbf{T}ransparency indicates visibility into microarchitectural or gate-level behavior—both crucial for all three key challenges. \textbf{P}rogrammability captures flexibility for post-fabrication analysis, enabling A/B testing and handling obfuscated hardware. \textbf{A}ccessibility reflects ease of use for hardware/software developers, especially in-field, supporting A/B testing and opaque software contexts.

\newcommand{\no}{\textcolor{red}{N}}

\begin{table}[tbp]


\centering
\begin{tabular}{p{0.35cm}|p{4.9cm}|p{0.2cm}|p{0.2cm}|p{0.2cm}|p{0.2cm}} 
\textbf{Yr}&\textbf{Technique} & \textbf{S} & \textbf{P} &  \textbf{T} & \textbf{A} \\ \hline                     
\multicolumn{6}{c}{General Approaches} \\ \hline
-- & SW profiling   & \no{}   & \no{}           & \no{}  & Y      \\
-- & Perf counters  & Y       & \no{}         & $p$        & Y      \\
-- & Debug monitors & Y       & \no{}     & Y        & \no{}       \\
-- & HW emulation     & \no{}    & Y           & Y     & \no{}      \\
-- & RTL simulation   & \no{}    & Y           & Y     & \no{}       \\
        \hline
\multicolumn{6}{c}{Security} \\ \hline
'03 & DISE \cite{1207014}        & \no{} & $p$  & \no{}& \no{} \\
'10 & FlexCore \cite{5695532}    & Y & $p$  &  Y & \no{} \\
'11 & LBA \cite{10.1145/1945023.1945034, 10.1109/MM.2009.6}    & \no{} & Y & \no{} & \no{} \\
'20 & PHMon/Nile \cite{244034, 8219379}  & Y & $p$ & \no{} & \no{} \\
\hline
\multicolumn{6}{c}{Embedded Systems} \\ \hline
'10\textsuperscript{1} & ABACUS\textsuperscript{3} \cite{matthews2010configurable, shannon2015performance, doyle2017performance}     & Y & $p$ & \no{} & Y\\ 
'13 & hidICE Verification \cite{backasch2013runtime}     & Y & Y & \no{} & \no{} \\
'15 & SOF \cite{lee2012event, lee2015system}        & Y & \no{} & $p$ & \no{} \\
'16\textsuperscript{1} & AIPHS\textsuperscript{2} \cite{moro2015hardware, valente2016flexible, valente2021composable}      & Y & \no{} & $p$  & Y  \\
'17 & Enhanced PMU\textsuperscript{2} \cite{scheipel2017system}           & Y & \no{} & $p$ & Y\\
'18 & NIRM \cite{seo2018non}       & Y & \no{} & \no{} & \no{} \\
\hline
\multicolumn{6}{c}{Profiling and PerfMonitors} \\ \hline
'01 & Programmable Co-Proc\cite{903267, 10.1145/1168917.1168890}   & Y & Y & $p$ & \no{} \\ 
'01 & Stratified Sampling \cite{sastry2001rapid} & \no{} & Y & $p$ & \no{}\\
'03 & ULF \cite{zhang2003unified}                    & $p$ & Y & \no{}  & \no{} \\ 
'03 & Interval Based Profiling \cite{narayanasamy2003catching} & \no{} & Y & $p$ & \no{}  \\
'05 & Owl \cite{10.1145/1062261.1062284}        & \no{} & $p$ & Y & Y \\
 \hline
'25 & \textbf{IPU (ours)}        & Y  & Y             & Y          & Y\\
 \hline
\multicolumn{6}{p{8.0cm}}{\textsuperscript{1}Year of most relevant work
\textsuperscript{2}Limited to manipulating event counts
\textsuperscript{3}Limited to architectural traces. $p$ means partial.}
\end{tabular}

\caption{Related work in our 4-axes taxonomy.
S (Speed); P (Programmability); A (Accessibility); T (HW Transparency)}
\label{tab:primitives-other-soln}
\label{tab:primitives-problems}

\end{table}

\textit{Software profiling} tools like \texttt{strace}, \texttt{gprof}, and binary instrumentation are well-established, but lack hardware transparency and suffer slowdown as introspection complexity increases (e.g., instruction count is fast with Pin, memory tracing is not). \textit{Performance counters} offer speed and accessibility and are the current state of the art, yet suffer from limited programmability and visibility—restricted to pre-defined events, with no support for A/B testing or richer analytics. Techniques such as PGO driven by perf counters~\cite{10444807, chen2010taming, chen2016autofdo, 36575, wicht2014hardware, novillo2014samplepgo, ramasamy2008feedback, dean1997profileme, merten1999hardware, liu2016sample, levin2008complementing} and Intel PT~\cite{intelPT} improve program optimization, but address only part of the hardware opacity problem.
\textit{Debug monitors}~\cite{jtag}, including JTAG and boundary-scan, offer deeper hardware access but require physical connections and cannot capture live software execution efficiently—making them inaccessible for typical users. \textit{HW emulation} platforms like Cadence Palladium~\cite{palladium} and Synopsys Zebu~\cite{zebu} offer full hardware transparency but are limited to chip designers, cost millions of dollars, and run at 2–5$\times$ slowdown. RTL simulation lowers cost but is orders-of-magnitude slower and equally inaccessible. \textit{Cycle-level simulators} improve speed slightly, but still suffer 100$\times$ or more slowdown, making them impractical for in-field use.
\revision{Table~\ref{tab:ipu-compare} compares IPU to two-SOTA introspection approaches.}

\begin{table}[tbp]
\small
\centering
\begin{tabular}{l|c|c|c}
\toprule
\textbf{Feature} & \textbf{CoreSight} & \textbf{Intel PT/PMU} & \textbf{IPU } \\
\midrule
Granularity        & Trace-level     & Event/sample     & Per-signal \\
Programmable       & No              & Configurable     & Full \\
Control interface  & Hardwired       & MSRs             & API  \\
\bottomrule
\end{tabular}
\caption{IPU compared to existing introspection mechanisms}
\label{tab:ipu-compare}
\vspace{-10pt}
\end{table}

The related academic works fall under 3 categories. Along the security angle, DISE \cite{1207014}, FlexCore \cite{5695532}, LBA\cite{10.1145/1945023.1945034, 10.1109/MM.2009.6}, and PHMon/Nile \cite{244034, 8219379}
focus on triggers based on events occurring and often run at-speed, however eschewing programmability, accessibility, and transparency. Driven by JTAG-like approaches, techniques from embedded systems, ABACUS\cite{matthews2010configurable, shannon2015performance, doyle2017performance},
hidICE Verification \cite{backasch2013runtime},
SOF \cite{lee2012event, lee2015system},
AIPHS \cite{moro2015hardware, valente2016flexible, valente2021composable}
Enhanced PMU \cite{scheipel2017system}, NIRM \cite{seo2018non},  allow at-speed analysis, in general assuming direct access to the hardware signals - thus lacking accessibility and programmability. Finally, novel uses of performance counters and profiling have been proposed, which suffer from lack of hardware transparency\cite{sastry2001rapid,zhang2003unified,narayanasamy2003catching,10.1145/1062261.1062284,west2009core,8574573}.

\section{Conclusion}\label{sec:conc}
This paper presents a detailed treatment of hardware introspection resolving three key challenges: Lack of A/B Testing for hardware, obfuscated Hardware, and obfuscated Software . We demonstrate an IPU can execute introspection programs in the field at real time speeds. This paper's key contributions are the definitions, system design, and implementation of the introspection processing unit (IPU) down to the level of an RTL implementation, including a comprehensive simulation testbed comprising both CPU and GPU designs. We show an implemented IPU adds $<1\%$ to area and power. The IPU provides unprecedented insights to hardware designers and software developers. 


\bibliographystyle{IEEEtranS}
\bibliography{references,references2,references3}

\end{document}